\documentclass{article}
\usepackage{amsmath}
\usepackage{graphicx} % For including images
\usepackage{caption} % For better caption formatting
\usepackage{authblk}
\usepackage{color,soul}
\usepackage{array}
\usepackage{booktabs}
\usepackage{multirow}
\usepackage{hyperref}
\usepackage{subcaption}
\usepackage{gensymb} 

\usepackage[margin=1in]{geometry}
\usepackage{amssymb}
\usepackage{xcolor}

\begin{document}
	
	\title{A Benchmark Dataset for Machine Learning Surrogates of Pore-Scale CO\(_2\)-Water Interaction}
	\date{}
	
	\author[1]{\small Alhasan Abdellatif$^ *$ }
	\author[1]{\small Hannah P. Menke$^ *$ }
	\author[1]{\small Julien Maes}
	\author[1]{\small Ahmed H. Elsheikh}
	\author[1]{\small Florian Doster}

	\affil[1]{\footnotesize Institute of GeoEnergy Engineering (IGE), School of Energy, Geoscience, Infrastructure \& Society, Heriot-Watt University, Edinburgh, EH14 4AS, UK}

	\maketitle
	\def\thefootnote{*}\footnotetext{These authors contributed equally to this work}
	\def\thefootnote{}\footnotetext{Corresponding authors: aa2448@hw.ac.uk and h.menke@hw.ac.uk}
	\begin{abstract}
		Accurately capturing the complex interaction between CO\(_2\) and water in porous media at the pore scale is essential for various geoscience applications, including carbon capture and storage (CCS). We introduce a comprehensive dataset generated from high-fidelity numerical simulations to capture the intricate interaction between CO\(_2\) and water at the pore scale. The dataset consists of 624 2D samples, each of size $512 \times 512$ with a resolution of $35 \mu$m, covering 100 time steps under a constant CO\(_2\) injection rate. It includes various levels of heterogeneity, represented by different grain sizes with random variation in spacing, offering a robust testbed for developing predictive models. This dataset provides high-resolution temporal and spatial information crucial for benchmarking machine learning models. 
	\end{abstract}
	
	\section*{Background \& Summary}
	% Importance and challenges of CO2 Transport in Porous Media
	CO\(_2\) transport through porous media plays a critical role in both natural and engineered processes, including subsurface carbon sequestration \cite{xiao2009effects,guiltinan2021two}, enhanced oil recovery \cite{xu2020pore}, and groundwater management \cite{cassiraga2005performance}.
	The challenge lies in accurately characterizing the movement and saturation of CO\(_2\), which is influenced by the complex interactions between fluid phases and the geological heterogeneity of the porous structure \cite{dentz2011mixing}. As CO\(_2\) is injected into underground formations, its movement through the pore spaces of geological materials, such as sandstone or basaltic reservoirs, dictates how efficiently it can be stored over long periods. This transport process is influenced by various factors, including capillary forces and chemical interactions between CO\(_2\), brine, and the mineral matrix. 
	
	%Methods for Estimating Transport Properties
	Various approaches are utilized to understand and predict CO\(_2\) transport in porous media. Laboratory techniques, such as core flooding experiments \cite{mohammed2022investigating}, yield effective bulk properties like permeability and residual saturation. Advanced imaging methods, like X-ray micro-tomography \cite{huang2023investigation}, allow visualization of pore-scale phenomena but have limitations, especially for dynamic processes. Numerical simulations, including lattice Boltzmann \cite{gao2017reactive}, pore-network modeling \cite{xiong2016review}, and direct numerical simulation \cite{maes2022geochemfoam}, offer more precise estimations of the fluid properties, however at a significant computational cost.

	%Emerging Role of Machine Learning in Porous Media Analysis
	Machine learning (ML) models are emerging as valuable tools for predicting CO\(_2\) behavior in porous media, serving as efficient surrogates for computationally expensive simulations. Recent advancements highlight ML's potential to estimate properties, like pressure build-up and saturation levels, with impressive speed and accuracy \cite{zhu2018bayesian,zhong2019predicting,wang2021physics,wen2021ccsnet,wen2022u,wen2023real}. The principle of these models is to learn the relationship between inputs—such as physical properties of porous media and engineering parameters—and outputs, like spatial and temporal fluid changes. Once trained on a set of representative samples, these models can generalize to predict unseen patterns, such as new permeability fields or different injection scenarios, with considerable efficiency. 
	
	However, challenges remain in terms of having a sufficient and diverse dataset for training robust models that generalize well across various scenarios. For example, current datasets often remain constrained to relatively small scales, such as maximum mesh sizes of $256 \times 256$ \cite{wang2021ml,feng2021fast,wang2022pore,ko2023prediction,meng2023transformer,poels2024accelerating}, which limits the ability of these models to capture fine-grained patterns necessary for accurate predictions in complex formations. Another key limitation is that most datasets designed for machine learning models focus on predicting the final state (e.g., after the injection duration) rather than capturing intermediate states \cite{feng2021fast,wang2022pore,ko2023prediction}. This limitation restricts the ability of models to capture the dynamic evolution of processes over time, which is crucial for understanding CO\(_2\) transient behaviors in real-world geological scenarios.
	
	% our dataset
	In this paper, we introduce a high-resolution dataset designed for benchmarking machine learning models in predicting CO\(_2\) behavior during multiphase flow in porous media. The dataset comprises 624 two-dimensional samples, each of size $512 \times 512$ pixels with a spatial resolution of $35 \mu$m, capturing the intricate interplay between CO\(_2\) and water over 100 equally spaced temporal snapshots under a constant CO\(_2\) injection rate. A distinctive feature of this dataset is its incorporation of varying levels of heterogeneity, represented through different grain sizes, which simulate realistic geological variability. This comprehensive dataset offers critical temporal and spatial granularity, serving as a utility for developing and benchmarking machine learning models.

	\section*{Methods}
	\subsection*{Geometry Preprocessing}

The pore structures are generated with the open‑source notebook
\texttt{DrawMicromodels.ipynb} \footnote{\url{https://github.com/hannahmenke/DrawMicromodels}, commit 5e0f947}
, which perturbs a regular triangular lattice of mean grain radius \(R_{0}\) by three \textit{heterogeneity amplitudes} \(\{\texttt{raddevmax},\texttt{xdevmax},\texttt{ydevmax}\}\).
For the \(n\)-th grain

\[
\begin{aligned}
  R_n &= R_0\bigl(1+\delta^{(R)}_n\bigr),\\[2pt]
  x_n &= x_n^{\text{lattice}} + L_x\,\delta^{(x)}_n,\\[2pt]
  y_n &= y_n^{\text{lattice}} + L_y\,\delta^{(y)}_n,
\end{aligned}
\]
where each perturbation term \(\delta\in[-a,a]\) is sampled from a uniform distribution whose half‑width \(a\) is the level‑dependent deviation listed in Table~\ref{tab:levels}.  Five levels are defined, ranging from well‑sorted media (Level 1) to highly heterogeneous media (Level 5).

\begin{table}[h]
  \centering
  \caption{Quantitative definition of the five heterogeneity levels
           (dimensionless amplitudes).}
  \label{tab:levels}
  \begin{tabular}{lccccc}
    \toprule
    \textbf{Level} & 1 & 2 & 3 & 4 & 5\\
    \midrule
    \(\texttt{raddevmax}\) & 0.05 & 0.10 & 0.15 & 0.20 & 0.25\\
    \(\texttt{xdevmax}\)   & 0.02 & 0.04 & 0.06 & 0.08 & 0.10\\
    \(\texttt{ydevmax}\)   & 0.02 & 0.04 & 0.06 & 0.08 & 0.10\\
    \bottomrule
  \end{tabular}
\end{table}

\paragraph{Physical motivation.}
The radius variation mimics sedimentary sorting, while positional jitter reproduces local compaction and packing irregularities observed in outcrop sandstones (\(C_{\!V}\approx0.05\)–0.25).  Increasing these amplitudes therefore widens the pore‑throat distribution and the capillary contrast, both of which are known to control CO$_2$–water displacement dynamics.

\begin{figure}[h!]
		\centering
		\includegraphics[width=\textwidth]{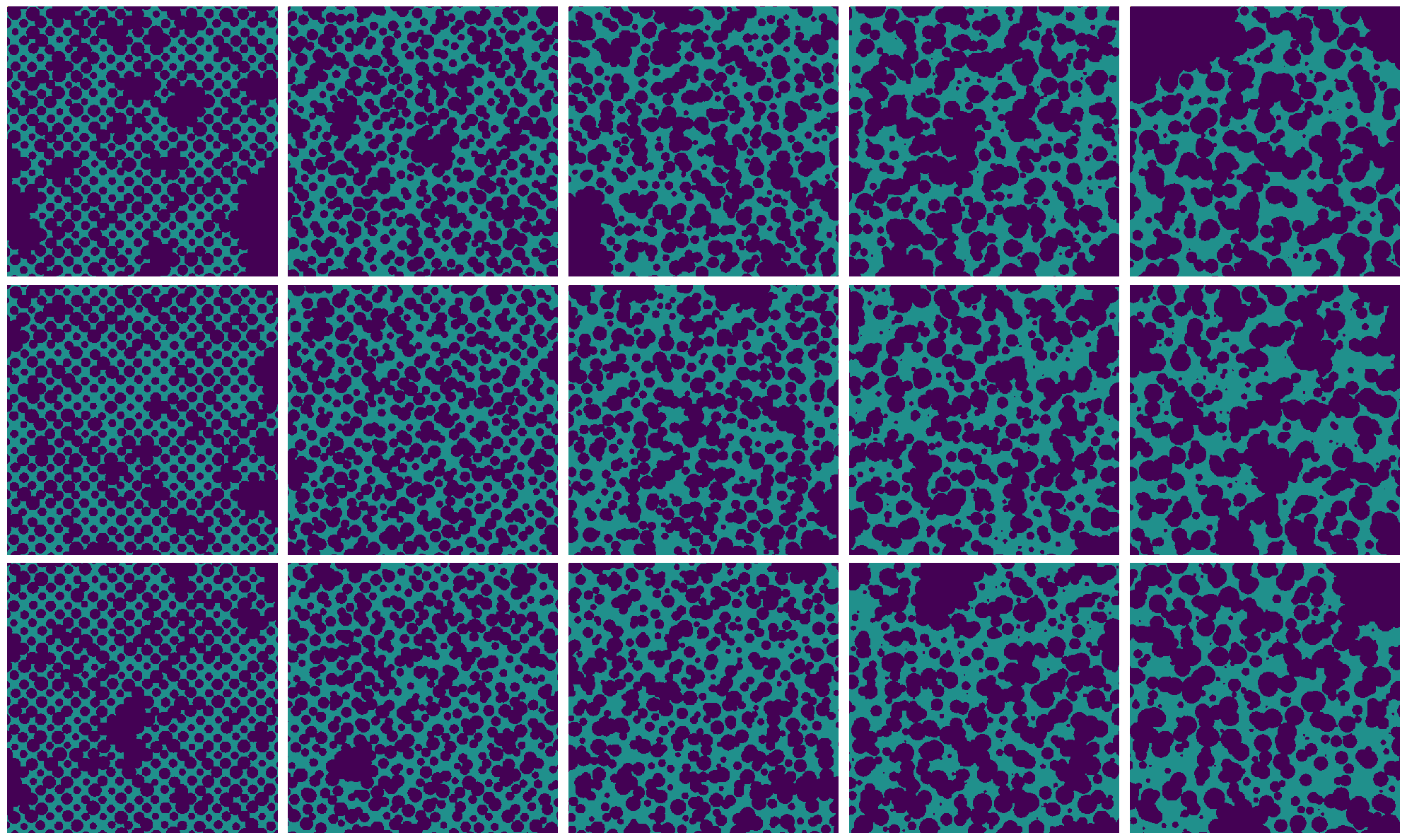} 
		\caption{Some examples of domain geometries corresponding to different patterns of heterogeneity. The heterogeneity level increases from left to right. }
		\label{fig:heterogeneity_samples}
	\end{figure}

\paragraph{Parametric sweep and augmentation.}
For each level we perform a deterministic sweep over
$R_0\in\{70,80,90\}$ and target porosities
$\phi\in\{0.20,0.25,0.30,0.35,0.40,0.45\}$, producing
\(5\times3\times6=90\) base images. Twelve images that displayed percolation shortcuts were discarded after visual inspection, leaving 78 accepted bases. Each \(1024\times1024\) image is subsequently cropped into four non‑overlapping quadrants (\(512\times512\)), and mirrored vertically.
This yields the final ensemble of \(78\times4\times2=624\)
geometries used in this study as shown in Figure \ref{fig:heterogeneity_samples}. By exposing the ML models to a range of grain size distributions and spatial configurations, the dataset enhances the model's ability to generalize to unseen porous media.%For instance, models trained on this dataset can learn to recognize patterns of heterogeneity that influence fluid dynamics, which is critical in applications such as CO\(_2\) sequestration, enhanced oil recovery, and groundwater flow modeling.
	
The inter‑sample sweep forces machine‑learning surrogates to learn scale‑invariant descriptors, while the intra‑sample jitter trains them to handle local anomalies, both are crucial for robust generalisation to unseen geological settings. It allows the ML model to develop robust feature extraction capabilities that are invariant to changes in grain sizes and configurations. This is crucial for ensuring that the predictions remain accurate across different geological formations. The dataset contains 624 geometries, each one is of size $512\times512$ and the physical resolution per pixel is $35 \mu$m. All samples are available in HDF5 format along with the simulations.
	
	\subsection*{Multi-phase flow at the pore-scale}
	% Multi-phase flow
	Understanding CO\(_2\) injection into water-filled porous media at the pore scale is critical for designing effective carbon storage strategies, especially in tight reservoirs where pore structures are highly heterogeneous and capillary forces dominate. At this scale, the interplay between fluid properties, pore geometry, and interfacial dynamics significantly influences the distribution and transport of CO\(_2\). These micro-scale interactions can lead to complex displacement patterns including snap off, coalescence, and ganglion migration that are difficult or impossible to capture with conventional Darcy-scale constitutive functions such as saturation-dependent capillary pressure and relative permeabilities. Robust Darcy-scale models however are key to predicting CO\(_2\) migration and storage efficiency.
	
	The two-phase flow simulations in this study were conducted using GeoChemFoam \cite{maes2022geochemfoam}, an advanced open-source numerical simulator developed at the Institute of GeoEnergy Engineering at Heriot-Watt University. GeoChemFoam is based on the OpenFOAM framework and is specifically designed to investigate pore-scale processes critical to energy transition and carbon storage.
	
	GeoChemFoam uses the algebraic Volume-of-Fluid method \cite{rusche2002computational} to solve multiphase flow. The velocity $\mathbf{u}$ and the pressure $p$ solve the single-field Navier-Stokes Equations (NSE):
	
	\begin{align}
	\nabla \cdot \mathbf{u} &= 0 \, , \\
	\rho \left(\frac{\partial u}{\partial t} + u \cdot \nabla u \right) &= -\nabla p + \nabla \cdot S + f_{st},
\end{align}
	where:
	\begin{itemize}
		\item $\rho = \alpha_{1}\rho_{1}+\alpha_{2}\rho_{2}$ is the fluid density,
		\item $\mathbf{u}$ is the velocity,
		\item $S$ = $\mu \left( \nabla u + \nabla u^{T}\right)$ is the viscous stress, 
		\item $\mu=\alpha_{1}\mu_{1}+\alpha_{2}\mu_{2}$ is the fluid viscosity,
		\item $p$ is the pressure,
		\item $f_{st}$ is the surface tension force, 
		\item $\alpha_{i}$ is the phase volume fraction, and
		\item $i=1,2$ refers to the phase index.
	\end{itemize}
	
	The surface tension force is approximated using the Continuous Surface Force (CSF) model \cite{rusche2002computational}:
	
	\begin{equation}
		f_{st} = \sigma \kappa \nabla \alpha_{1},
	\end{equation}
	
	where:
	\begin{itemize}
		\item $\sigma$ is the interfacial tension, and
		\item $\kappa = \nabla \cdot \left(\frac{\nabla \alpha_{1}}{|\nabla \alpha_{1}|}\right)$ is the interface curvature.
	\end{itemize}
	
	The phase indicator function $\alpha_{1}$ solves the phase transport equation:
	
	\begin{equation}
		\frac{\partial \alpha_{1}}{\partial t} + \nabla \cdot (\alpha_{1} u) + \nabla\cdot \left(\alpha_{1}\alpha_{2} u_{r} \right)= 0.
	\end{equation}
	
	To reduce interface smearing, an artificial compression term is introduced by replacing $u_{r}$ with a compressive velocity \cite{rusche2002computational}.

	\begin{figure}[h!]
		\centering
		\includegraphics[width=0.5\textwidth]{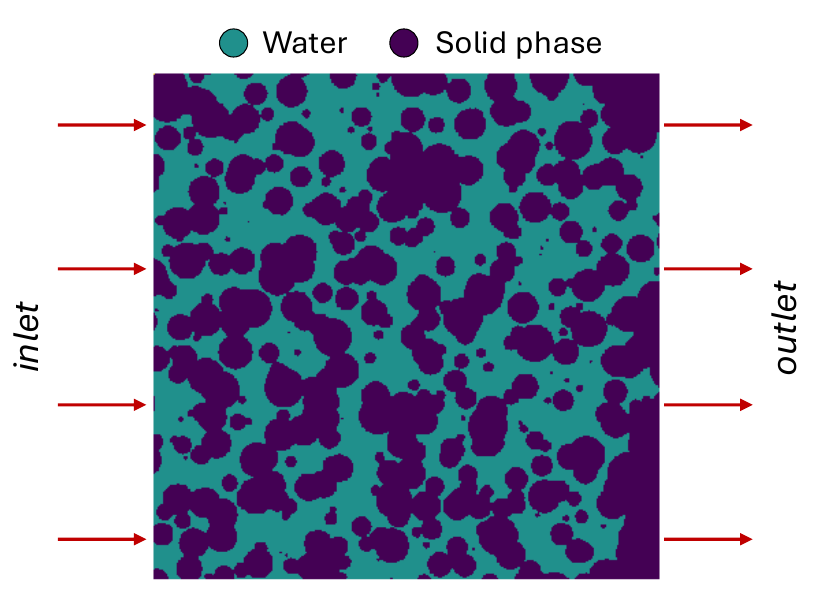} 
		\caption{Visualization of CO\(_2\) injection in porous media initially saturated with water. The CO\(_2\) is injected from the left boundary, displacing the water phase as it migrates through the pore space.}
		\label{fig:simulation_diagram}
	\end{figure}
	
	Each geometry is a domain of 512 $\times$ 512 voxels at a resolution and depth of 35 microns. We perform a two-phase flow simulation where CO\(_2\) is injected into a fully water-filled model from the left boundary, as shown in Figure \ref{fig:simulation_diagram}, at a flow rate of $1 \times 10^{-8} m^3/s$ corresponding to a capillary number of approximately $5 \times 10^{-6}$. The CO\(_2\) properties are set to be $\mu_{CO_{2}} = 7.37 \times 10^{-8} m^2/s$ and $\rho_{CO_{2}} = 3.84 \times 10^{2} kg/m^3$. The water properties are $\rho_{water} = 1 \times 10^{3} kg/m^3$ and $\mu_{water} = 1 \times 10^{-6} m^2/s$, with the interfacial tension between phases at 0.03 $N/m$, and the contact angle $\theta = 45\degree$. The simulation was run until a total time of 1 s with a write interval of 0.01 s and a convergence tolerance of $1 \times 10^{-8}$. 
	
\begin{figure}[h!]
		\centering
		\includegraphics[width=1\textwidth]{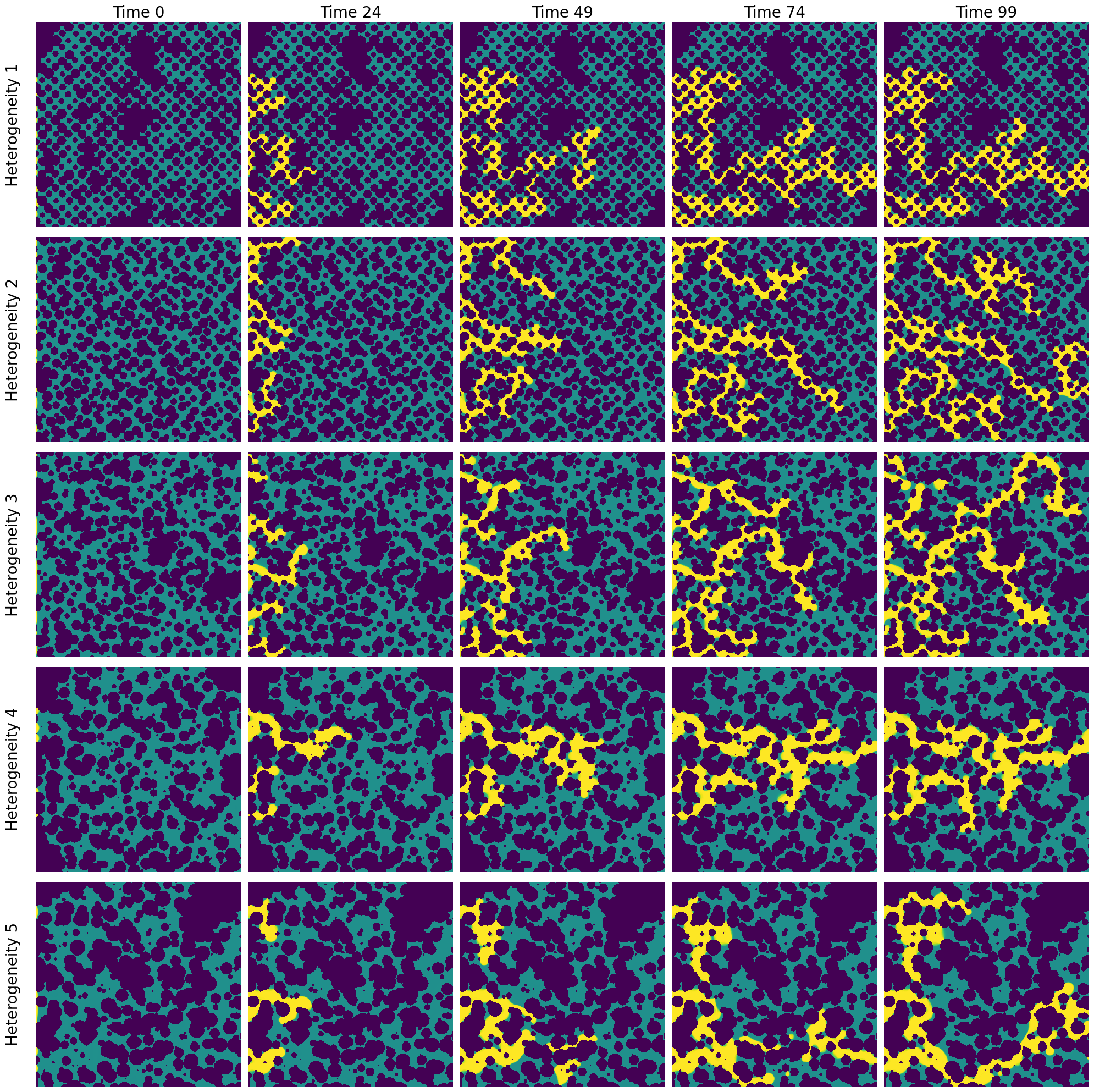} 
		\caption{CO\(_2\) (yellow) displacing water in a porous media during the simulation time. Each row shows an example of the 5 heterogeneity levels in the dataset.}
		\label{fig:displacement}
	\end{figure}

\begin{figure}[h!]
		\centering
		\includegraphics[width=1\textwidth]{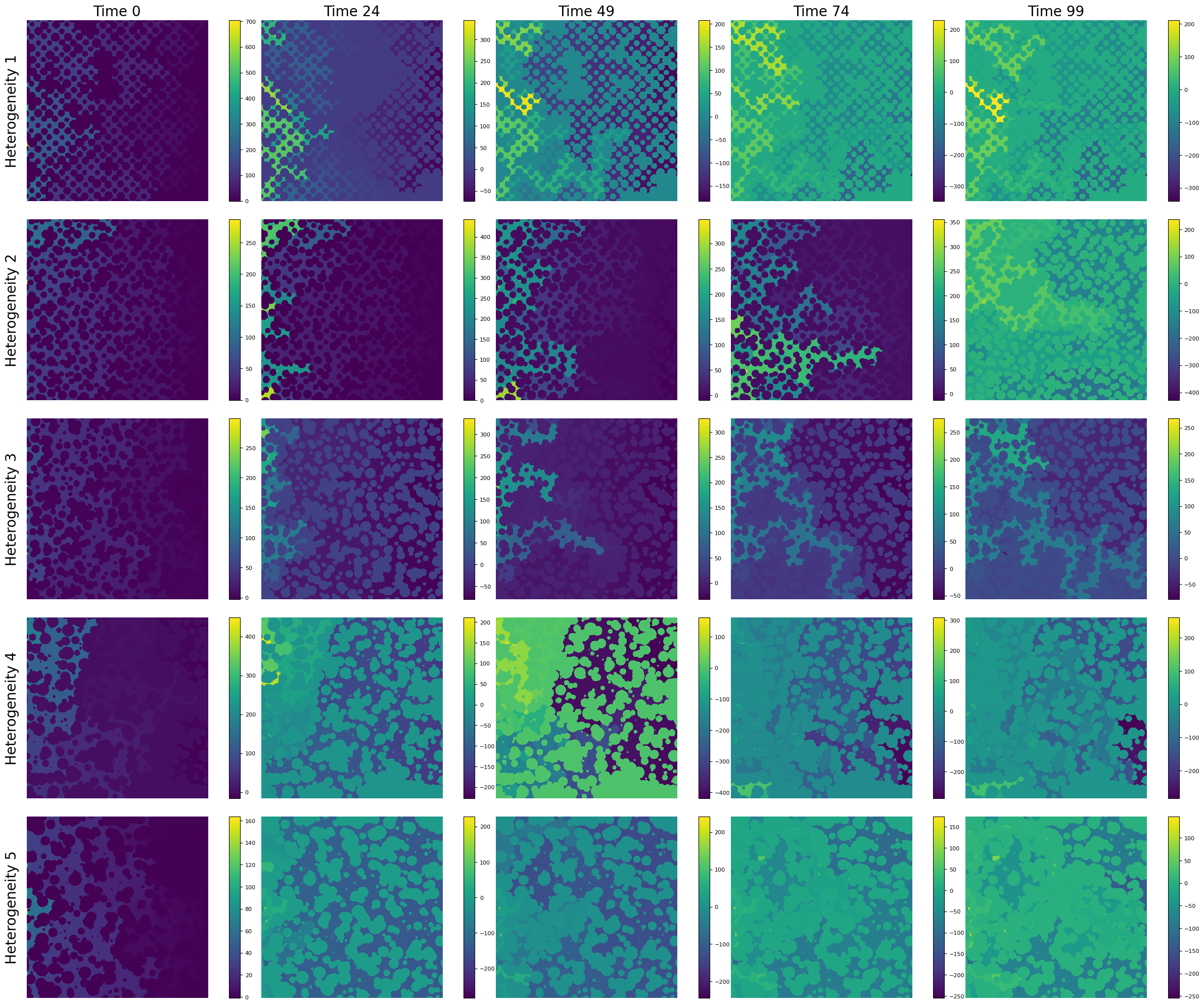} 
		\caption{Pressure field at different injection duration. Each row shows an example of the 5 heterogeneity levels in the dataset.}
		\label{fig:Pressure_fields}
	\end{figure}

\begin{figure}[h!]
		\centering
		\includegraphics[width=1\textwidth]{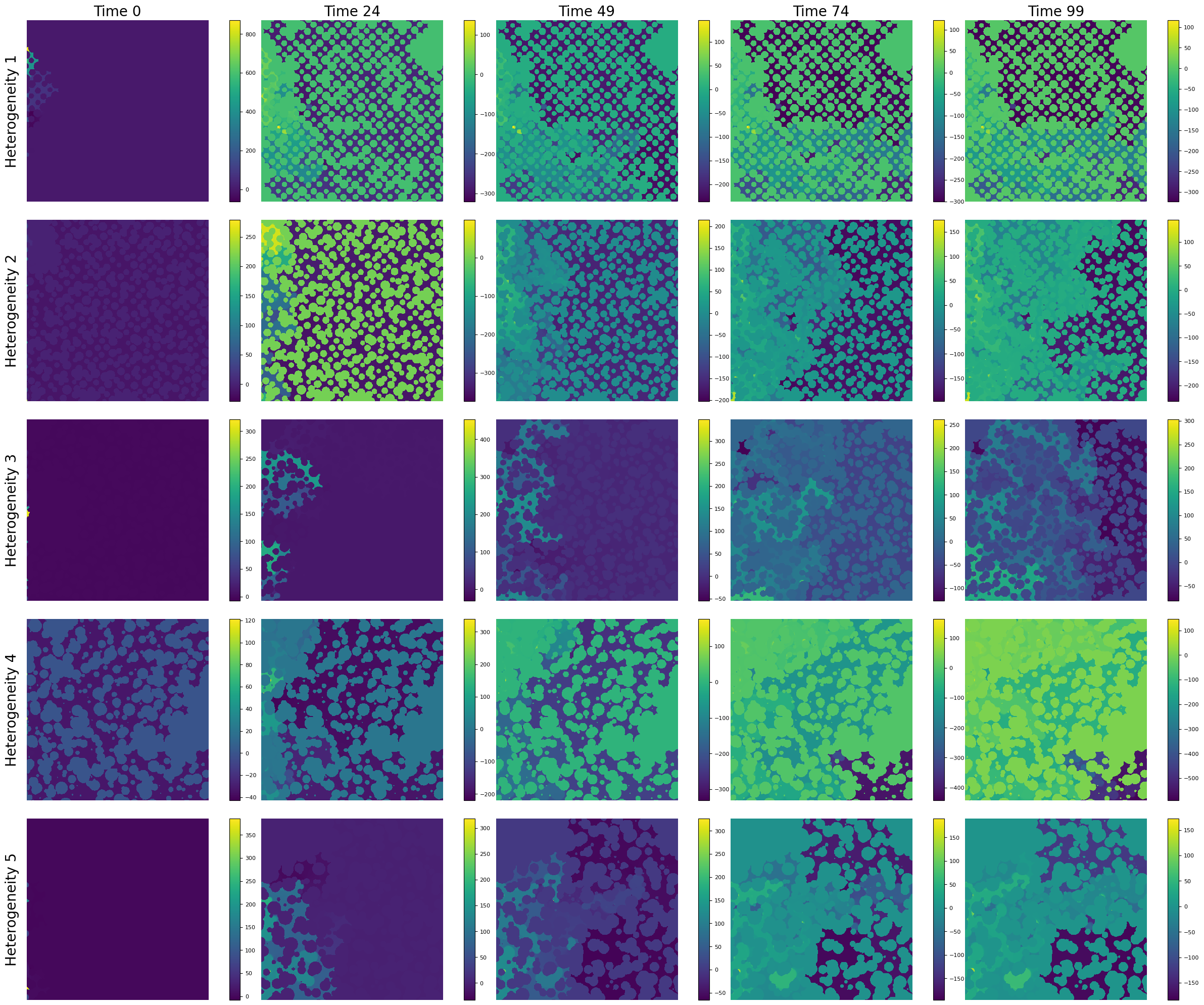} 
		\caption{Capillary pressure field at different injection duration. Each row shows an example of the 5 heterogeneity levels in the dataset.}
		\label{fig:pc_fields}
	\end{figure}

\begin{figure}[h!]
		\centering
		\includegraphics[width=1\textwidth]{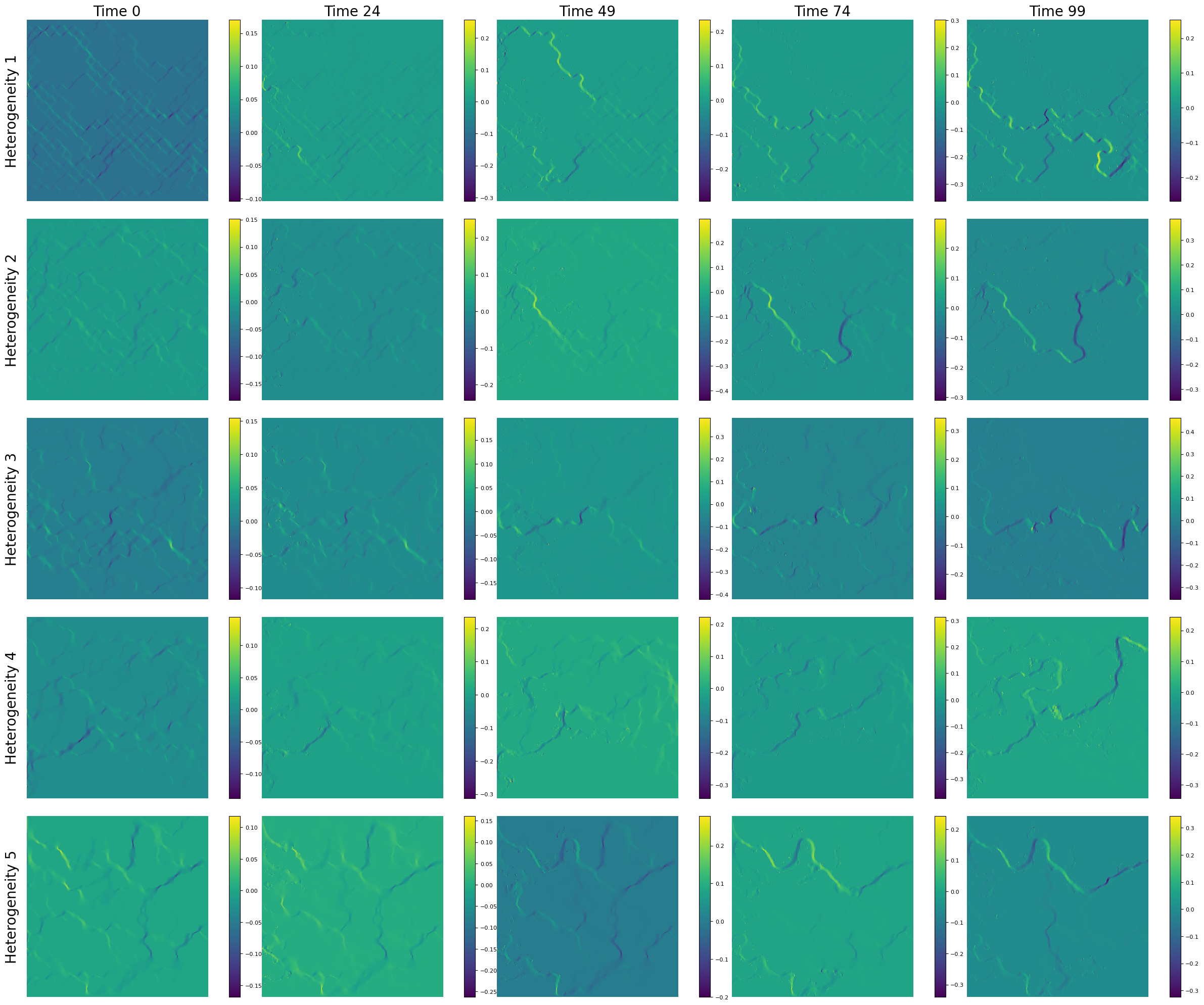} 
		\caption{Vertical velocity field at different injection duration. Each row shows an example of the 5 heterogeneity levels in the dataset.}
		\label{fig:Vertical_velocity_Field}
	\end{figure}

	In Figure \ref{fig:displacement}, we show the CO\(_2\) migration pattern, for different heterogeneities, as it displaces water at different time steps. Over time, the CO\(_2\) saturation front expands, displaying distinct channelized patterns and regions of accumulation. These patterns demonstrate the interaction between capillary forces, viscous forces, and the underlying geological features. The time-lapse progression also reveals the impact of grain size and pore structure on flow dynamics, emphasizing the importance of micro-scale processes in controlling large-scale behavior. We also show the pressure, capillary pressure, and vertical velocity fields for different geometries in Figures \ref{fig:Pressure_fields}, \ref{fig:pc_fields}, and \ref{fig:Vertical_velocity_Field}, respectively.
	
	\section*{Data Records}
   The dataset has been made available on \url{https://doi.org/10.5061/dryad.jm63xsjn5} \cite{abdellatif2025benchmark} and is organized into 10 folders, with each of the 5 geometries having its original version and a vertically flipped version ($2 \times 5 = 10$). The simulation samples are provided in HDF5 format, with each file including water saturation (\(\alpha_{water}\)), pressure (\(p\)), capillary pressure (\(pc\)), horizontal velocity (\(U_x\)), vertical velocity (\(U_y\)), and a binary image of the physical domain (where pores are denoted by 1 and grains by 0), as detailed in Table~\ref{tab:hdf5_files} which also lists the keys required to access the data. The water saturation \(\alpha_{water}\) is in the range \([0,1]\); hence, the CO\(_2\) saturation field can be computed using the relation \(\alpha_{CO_2} = (1-\alpha_{water}) \times \text{img}\), where \(\text{img}\) denotes the binary physical domain. Additionally, CSV files containing values for porosity, permeability, and relative permeability are provided, with details presented in Table~\ref{tab:csv_files}.

	\begin{table}[h!]
		\centering
		\begin{tabular}{|l|c|c|c|}
			\hline
			\textbf{File Name} & \textbf{Key} & \textbf{Size} & \textbf{Description} \\
			\hline
			\multirow{6}{*}{*.hdf5}
			& Ux & $100\times512^2$ & x-component of flow velocity \\  \cline{2-4}
			& Uy  & $100\times512^2$& y-component of flow velocity \\ \cline{2-4}
			& alpha\_water  & $100\times512^2$  & water saturation field over time \\ \cline{2-4}
			& img & $512^2$ & physical domain \\ \cline{2-4}
			& p &  $100\times512^2$ & pressure field \\ \cline{2-4}
			& pc &  $100\times512^2$  & capillary pressure field \\ 
			\hline
		\end{tabular}
		\caption{Overview of the dataset files, including flow velocity components, pressure fields, and physical domain representations with corresponding sizes and descriptions. Keys are provided for accessing hdf5 files.}
		\label{tab:hdf5_files}
	\end{table}

	\begin{table}[h!]
		%\centering
		\begin{tabular}{|l|p{10cm}|}
			\hline
			\textbf{File Name} & \textbf{Description} \\
			\hline
			poroPerm.csv & 
			Time, porosity, permeability $(m^2)$, the characteristic pore length L, 
			the Reynolds number Re, and the Darcy velocity $U_{D}$ at the beginning 
			of the simulation before any CO\(_2\) is injected into the model. \\
			\hline
			relperm.csv & 
			Porosity, permeability $(m^2)$, and the capillary 
			number of each phase (Ca$_1$ for water and Ca$_2$ for CO\(_2\)) at the 
			beginning of the simulation. The saturation of water $S_{w}$, the relative 
			permeability of water $k_{rw}$, and the relative permeability of CO\(_2\) 
			$k_{wo}$ are shown for each output timestep. \\
			\hline
		\end{tabular}
		\caption{List of files describing porosity and relative permeability values.}
		\label{tab:csv_files}
	\end{table}

	\section*{Technical Validation}
	
	The GeoChemFoam solver used for flow simulation has been validated against experimental data in \cite{zhao2019comprehensive}. For accurate approximation, a convergence tolerance of $1 \times 10^{-8}$ was used for all samples. 

	To assess the dataset's utility for improving model generalization, three models of a U-Net architecture \cite{ronneberger2015u} were trained on datasets of varying levels of heterogeneity. Each model was trained to predict future  CO\(_2\) saturation by mapping a sequence of four consecutive saturation maps to the subsequent four timesteps. During evaluation, these models were applied in an autoregressive fashion to generate long-term predictions up to 60 timesteps. Model A was trained on the full dataset (5-Levels), model B was trained on a subset containing four of the five levels (4-Levels), and model C was trained on a subset with only the first level (1-Level). All models were then evaluated on samples from the fifth level, unseen by models B and C. For this analysis, all input samples were resized to 256$\times$256 pixels, and predictions were made for the first 60 timesteps.

\begin{figure}[h!]
    \centering
    \includegraphics[width=\textwidth] {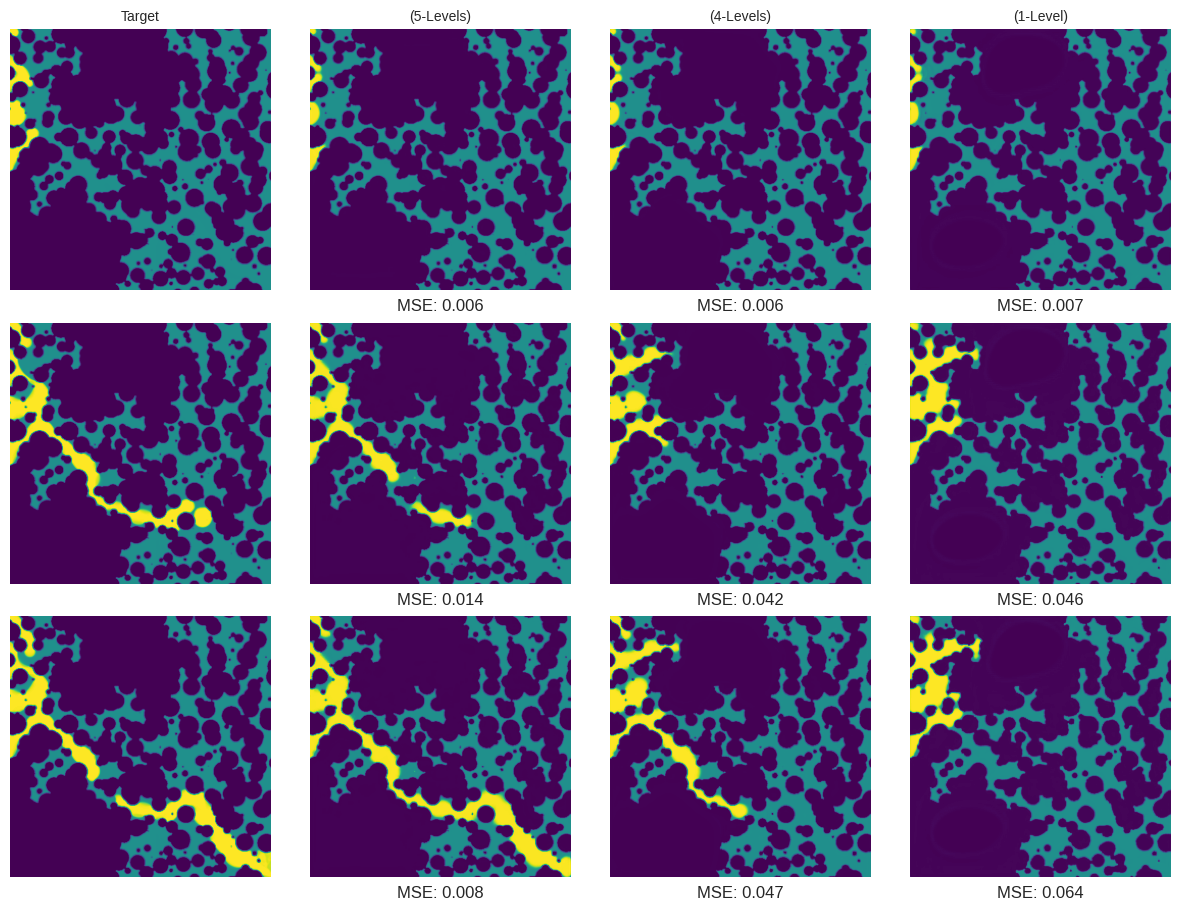}
    \caption{Qualitative comparison of model predictions against the target simulation for a sample from the test set.}
    \label{fig:sim_comparison}
    \end{figure}

\begin{figure}[h!]
     \centering
     \includegraphics[width=\textwidth]{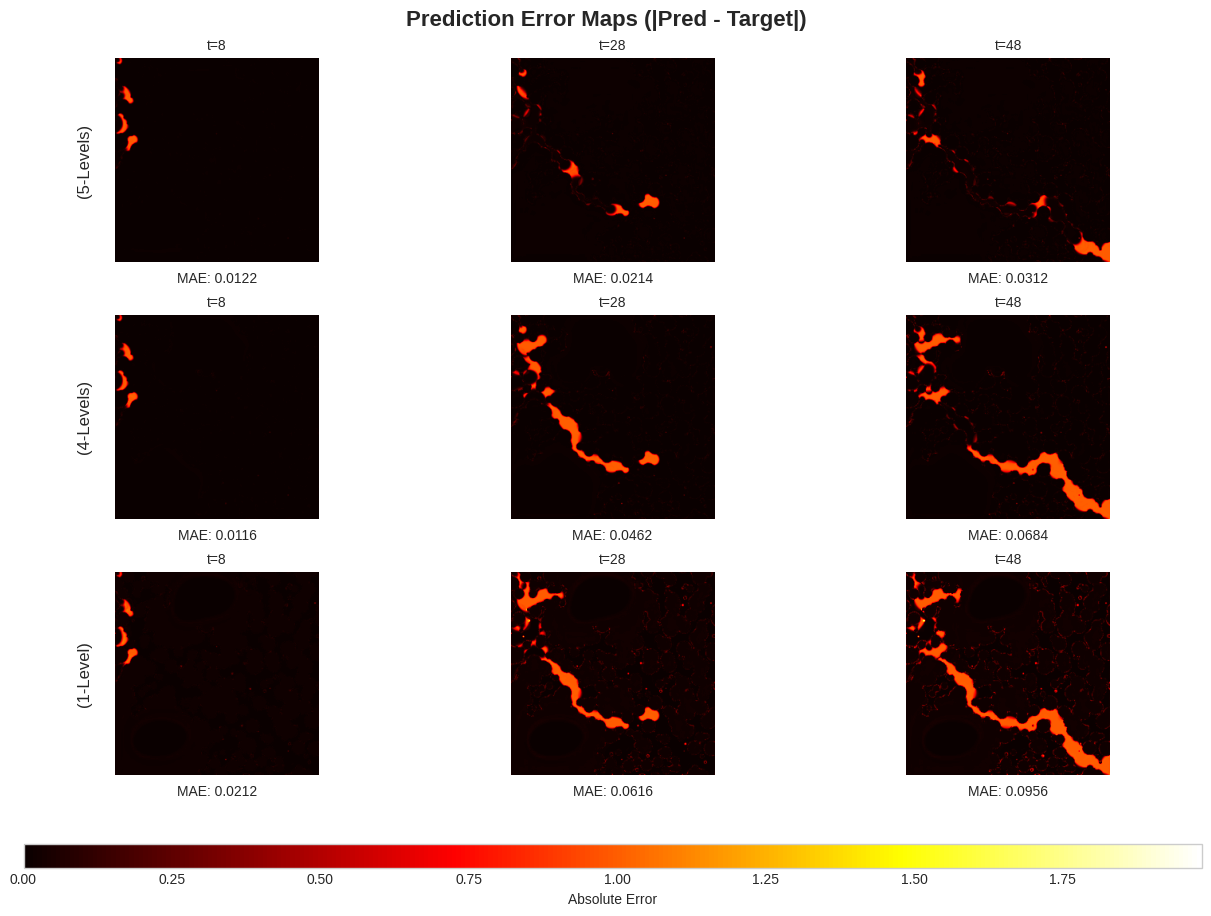}
     \caption{Prediction error maps for each model at different timesteps.}
     \label{fig:error_maps}
     \end{figure}

\begin{figure}[h!]
     \centering
     \includegraphics[width=\textwidth]{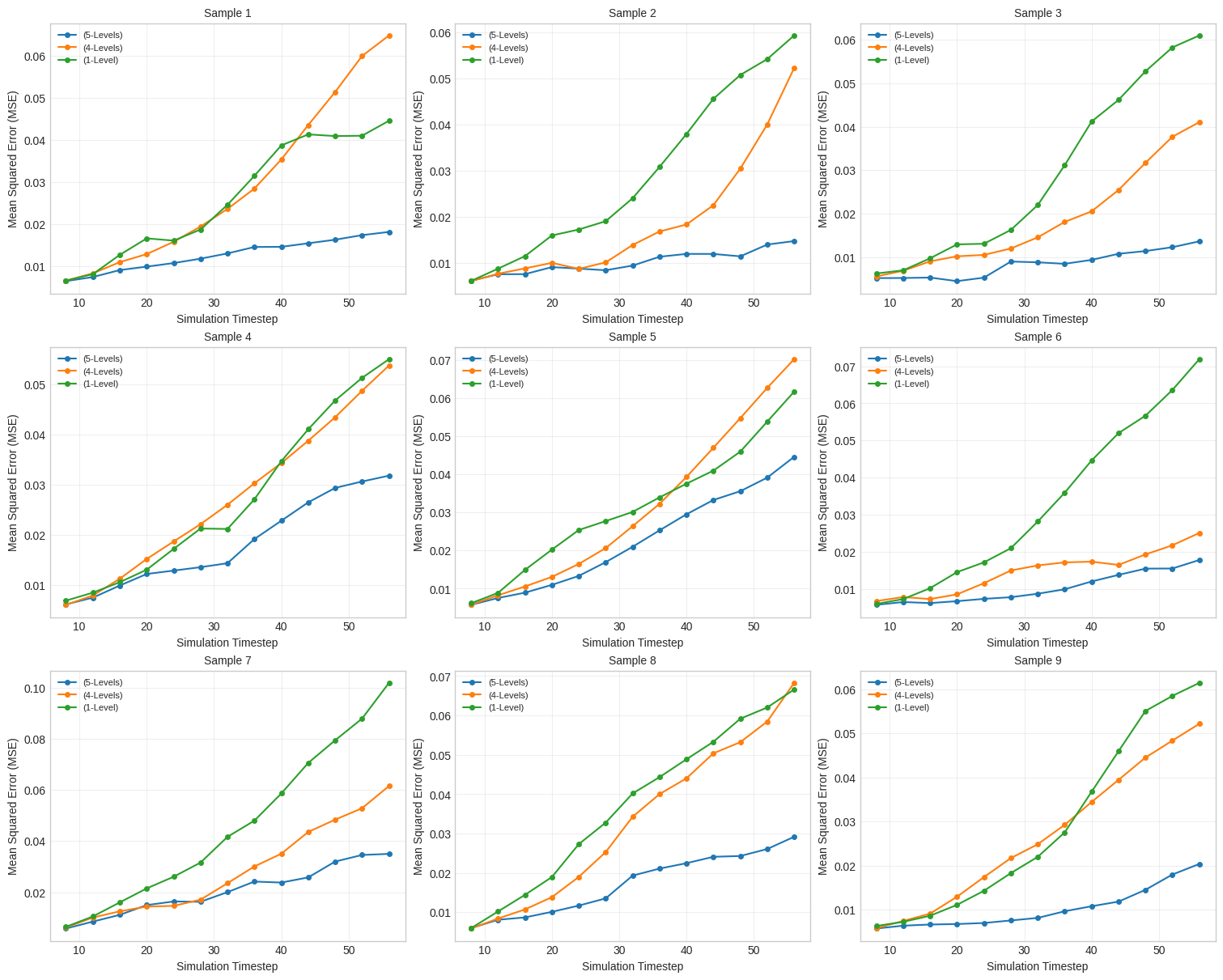}
     \caption{Mean Squared Error (MSE) over simulation timesteps for various samples of level 5.}
     \label{fig:mse_plots}
     \end{figure}

    The results, summarized in Table~\ref{tab:summary_stats}, indicate a clear benefit to training on a more diverse dataset. The 4-Levels model achieved a lower Mean Squared Error (MSE) on average (0.0254) across the test samples compared to the 1-Level model (0.0320). This demonstrates superior average performance and generalization. The 5-Levels model, having been trained on the test data, served as a benchmark and predictably achieved the lowest average MSE (0.0145). A direct visual comparison of the predicted simulations against the ground truth, as seen in Figure~\ref{fig:sim_comparison}, corroborates these quantitative findings. Furthermore, the qualitative error maps in Figure~\ref{fig:error_maps} visualize this trend, showing progressively lower absolute error from the 1-Level to the 5-Levels model. However, the per-sample MSE plots in Figure~\ref{fig:mse_plots} reveal that this improvement was not uniform across all samples; in some cases, the 4-Levels model performed similarly to, or slightly worse than, the 1-Level model. This suggests that while training on more varied data helps the model learn more general rules, it can also introduce biases that hinder performance on specific out-of-distribution samples. The primary conclusion is that increased training data diversity leads to better \textit{average} generalization, though not necessarily universal improvement on every individual sample.

      \begin{table}[h!]
        \centering
        \caption{Summary statistics for model performance on the unseen fifth level.}
        \label{tab:summary_stats}
        \begin{tabular}{lccc}
        \hline
        \textbf{Model Name} & \textbf{Mean MSE} & \textbf{Final Step MSE} & \textbf{Std Dev} \\
        \hline
        5-Levels & 0.014484 & 0.009853 & 0.004364 \\
        4-Levels & 0.025410 & 0.023486 & 0.007635 \\
        1-Level  & 0.032036 & 0.037971 & 0.008166 \\
        \hline
        \end{tabular}
    \end{table}

	\section*{Code Availability}
	The input files used to simulate CO\(_2\) flow is built using GeoChemFoam \cite{maes2022geochemfoam} and is available at \url{https://github.com/ai4netzero/generating_co2_flow}. The code is written in Python 3.11.9 and the list of the requirements is shown in the readme file. GeoChemFoam can be downloaded from \url{https://github.com/GeoChemFoam/GeoChemFoam-5.1} and has been validated against experimental data in \cite{zhao2019comprehensive}. 
	\section*{Acknowledgments}
        This work is funded by the Engineering and Physical Sciences Research Council's ECO-AI Project grant (reference number EP/Y006143/1), with additional financial support from the PETRONAS Centre of Excellence in Subsurface Engineering and Energy Transition (PACESET).
	\section*{Author contributions}
	Conceptualization and methodology, H.P.M., J.M., A.A., A.H.E.; visualization and writing, A.A., H.P.M.; formal analysis, A.A., H.P.M., A.H.E.; funding acquisition, A.H.E., F.D.,  H.P.M.; supervision, A.H.E, F.D., H.P.M. All authors have read and agreed to the published version of the manuscript.
	\section*{Competing interests}
	The authors declare no competing interests.

\end{document}